\documentclass{aa}  

\usepackage{graphicx}
\usepackage{txfonts}
\usepackage{epsfig}
\usepackage{xspace}
\usepackage{multicol}
\usepackage{multirow}
\usepackage{placeins}           

\usepackage{orcidlink}

\newcommand{\Ch}{\textit{Chandra}\xspace}
     
\newcommand{\Rsun}{\,R$_\odot$\xspace}    
\newcommand{\Msun}{\,M$_\odot$\xspace}     
\newcommand{\Msunyr}{\,M$_\odot$\,yr$^{-1}$\xspace}  
 
\newcommand{\kms}{\,km\,s$^{-1}$\xspace}

\newcommand{\mJybeam}{\,mJy\,beam$^{-1}$\xspace}     
\newcommand{\uJybeam}{\,$\mu$Jy\,beam$^{-1}$\xspace}

\begin{document} 

    \title{Resolving the heart of the Wolf-Rayet binary \textit{Apep} with ALMA}
    
    \author{S. del Palacio\orcidlink{0000-0002-5761-2417}\inst{1}\corrauth{santiago.delpalacio.astro@gmail.com}
            \and
            W. Vlemmings\inst{1}
            \and 
            M. Fern\'andez-L\'opez\orcidlink{0000-0001-5811-0454}\inst{2,3,4}
            \and 
            M. De Becker\inst{5} 
            \and
            P. Benaglia\inst{4}
            \and 
            V. Bosch-Ramon\inst{6}
            \and 
            B. Marcote\inst{7} 
            \and
            Luis A. Zapata\orcidlink{0000-0003-2343-7937}\inst{8}
            }
    
    \institute{Department of Physics and Astronomy, Chalmers University of Technology, SE-412 96 Gothenburg, Sweden
    \and 
    Institut de Ci\`encies de l’Espai (ICE-CSIC), Campus UAB, Carrer de Can Magrans S/N, E-08193 Cerdanyola del Valles, Catalonia, Spain
    \and
    Facultad de Ciencias Astron\'omicas y Geof\'isicas, Universidad Nacional de La Plata, Paseo del Bosque S/N, B1900FWA La Plata, Argentina 
    \and 
    Instituto Argentino de Radioastronom\'{\i}a (CCT-La Plata, CONICET; UNLP; CICPBA), C.C No 5. 1894, Villa Elisa, Argentina
    \and
    Space sciences, Technologies and Astrophysics Research unit -- STAR, University of Li\`ege, Quartier Agora, 19c, All\'ee du 6 Ao\^ut, B5c, B-4000 Sart Tilman, Belgium
    \and
    Departament de F\'{i}sica Qu\`antica i Astrof\'{i}sica, Institut de Ci\`encies del Cosmos (ICC), Universitat de Barcelona (IEEC-UB), Mart\'{i} i Franqu\`es 1, E08028 Barcelona, Spain
    \and
    Joint Institute for VLBI ERIC, Oude Hoogeveensedijk 4, 7991 PD Dwingeloo, the Netherlands
    \and
    Instituto de Radioastronom\'ia y Astrof\'isica, Universidad Nacional Autónoma de M\'exico, Morelia, Michoac\'an 58089, México
                }

   \date{Received ---; accepted ---}

  \abstract
   {The Wolf-Rayet (WR) phase is a critical stage in the evolution of massive stars. During this phase, the stars experience a large mass loss in the form of fast winds that are enriched with heavy elements. There are only a few WR binaries in the Galaxy, with \textit{Apep} being a truly exceptional system exhibiting a bright colliding-wind region at radio wavelengths and an impressive dust plume in the infrared. Although the angular separation between the two (WN and WC) components has been indirectly inferred to be about 47 mas, it has yet to be confirmed through direct measurements.} 
   {We aim to image the \textit{Apep} system at millimetre wavelengths to determine the position of the WR stars and their mass-loss rates.}
   {We observed \textit{Apep} with ALMA in Bands 6 and 7 in a large baseline configuration, achieving an angular resolution of $\approx$20~mas. We used a canonical spherical wind model to convert the measured flux densities into wind mass-loss rates.}
   {Towards \textit{Apep}, we detect point-like emission coming from the WN star and elongated emission most likely from the colliding-wind region; the WC star is not detected. We also detect the O star at 0.7$''$ that has been associated with the system. Our analysis favours mass-loss rates for the WN and WC components of $\dot{M}_\mathrm{WN}\sim 2\times10^{-5}$\Msunyr and $\dot{M}_\mathrm{WC}\sim10^{-5}$\Msunyr, respectively.} 
   {The mass-loss rates of the WR stars are significantly lower than previously reported, and we examine several factors that may account for this discrepancy. The diffuse emission from the colliding-wind region cannot be accounted for by the synchrotron flux, and thus additional free--free or dust emission is present.} 


   \maketitle
   \nolinenumbers
   
%

\section{Introduction}

The study of massive stars (those with initial masses $>8$~\Msun) is of key importance for several areas of astrophysics. Despite being only a minority in the overall stellar population, massive stars are major sources of ionising radiation and feedback to the interstellar medium (ISM). During their lifetime, they launch powerful and massive winds that continuously inject mechanical energy and chemically rich material into the ISM. In addition, the evolution of massive stars has major consequences, as they are the progenitors of supernova explosions and compact objects (neutron stars and black holes). In particular, observations of Wolf-Rayet (WR) stars are key to understanding the post-main-sequence evolution of massive stars \citep[e.g.][]{Langer2012}. 

The mass-loss rate of the winds, $\dot{M}$, plays a crucial role in the evolution and fate of massive stars \citep{Hirschi2015}, and even the dynamical evolution of supernova explosions as the ejecta interacts with the material previously expelled via stellar winds \cite[e.g.][]{vanMarle2010}. However, values of $\dot{M}$ are largely uncertain due to difficulties in their determination \citep{Puls2008}. In particular, very massive stars going through the WR phase are characterised by powerful winds with very high mass-loss rates of $\dot{M} \sim 10^{-6}$--$10^{-5}$~\Msunyr \citep{Cappa2004} and fast terminal velocities of $v_\infty = 1000$--$3000$~\kms, although again with very high uncertainties.

Massive stars are most likely found in interacting binary systems \citep{Sana2012, Offner2023}. In these systems, the powerful stellar winds collide, producing strong shocks in the colliding-wind region (CWR), which can accelerate relativistic particles that produce non-thermal emission \citep{Eichler1993}. The spectral energy distribution (SED) of a colliding-wind binary has different components that are relevant at radio to sub-mm wavelengths. Typically, the synchrotron emission from the CWR dominates at frequencies $<10$~GHz, as it has a negative spectral index\footnote{Here we use the convention that the flux density is $S_\nu \propto \nu^\alpha$.} \citep{DeBecker2013}. However, at lower frequencies ($<1$~GHz) this emission can be severely absorbed by the ionised stellar winds \citep[e.g.][]{Dougherty2003,delPalacio2022,Tasseroul2025}.
At higher frequencies ($>10$~GHz), the SED is dominated by thermal (free--free) emission. The stellar winds have a density profile that leads to an integrated spectrum that is between optically thick ($\alpha = 2$) and optically thin ($\alpha = -0.1$), with a canonical spectral index of $\alpha = 0.6$ \citep{Wright1975, Panagia1975}, although it can be slightly higher depending on the wind structure \citep{Leitherer1991} and velocity profile \citep{Daley-Yates2016}, with typical values of $\alpha \sim 0.76$ found in WR stars \citep{Nugis1998}. 
In addition, the ionised gas in the CWR can also produce thermal emission in the GHz--THz frequency range \citep{Pittard2006radio, Pittard2010radio}. Broadband measurements of the radio SED allow us to disentangle these emission components. 

The system \textit{Apep} (\object{2XMM~J160050.7$-$514245}), located at $RA = 10^\mathrm{h}\,43^\mathrm{m}\, 57.5^\mathrm{s}$, $DEC=-59\degr 32\arcmin 51.4\arcsec$ (J2000), was discovered by \cite{Callingham2019} and exhibits exceptional properties. First, it presents a spectacular dust plume seen at IR wavelengths, which introduces large extinction at optical frequencies. Second, it hosts the only confirmed binary with two WR stars (WN + WC) in the Galaxy \citep{Rosslowe2015}. 
Third, \textit{Apep} is a very powerful non-thermal emitter; it is the most luminous synchrotron emitter at radio frequencies among the known colliding-wind binaries and one of the few binaries with non-thermal emission detected at hard X-rays \citep{delPalacio2023}. Fourth, the binary has a long period $> 100$~yr \citep{Han2020, Han2025}, which allows us to resolve its components \citep{Marcote2021} and combine multi-epoch observations as the orbital evolution is currently very slow \citep{Bloot2022, White2025}. In addition, \cite{White2025} showed that the O-type star located $\approx 0.7"$ north is indeed a bound component of the \textit{Apep} system. 

One key piece of information in the study of evolved stars, and \textit{Apep} in particular, is the stellar wind properties. So far, several constraints and estimates of the mass-loss rates and terminal wind velocities have been presented based on spectroscopic analysis \citep{Callingham2020, Zhekov2025}, imaging \citep{Marcote2015, Han2025}, and SED modelling \citep{delPalacio2022}. One particular avenue of progress is the use of observations in the mm band, where the dominant emission component comes from the ionised stellar winds. In particular, this emission has a flux density $S_\nu \propto \dot{M}^{4/3}$ \citep{Wright1975}, such that measurements of the mm flux densities provide reliable estimates of the mass-loss rates depending only on simple assumptions about the stellar wind structure \citep[e.g.][]{Benaglia2015}. 

In this letter, we present a novel view of \textit{Apep} using high-resolution observations obtained with the Atacama Large Millimeter/submillimeter Array (ALMA), providing the first resolved imaging of the WR binary at mm wavelengths. We discuss the implications of these findings for the stellar winds in the context of other multi-frequency data of the system at lower angular resolution. 

\section{Observations and data reduction}\label{sec:observations}

\textit{Apep} was observed with the ALMA during Cycle 11 under project 2024.1.00060.S (PI: S.~del Palacio) using the 12-m array in its extended long-baseline configuration. The observations were obtained in Band~6 \citep[B6,][]{Ediss04} and Band~7 \citep[B7,][]{Kerr2004}, providing complementary high-angular-resolution continuum measurements. 

The B6 observations were carried out on 2025 July 23 and 2025 July 29, comprising two execution blocks with a total on-source integration time of 943.5~s. The correlator was configured in Frequency Division Mode (FDM) with four spectral windows centred at 229.00, 231.00, 242.00, and 244.00~GHz, each providing a bandwidth of 1.875~GHz, for a total continuum bandwidth of 7.5~GHz. Projected baselines ranged from 226~m to 15.24~km, resulting in a synthesised beam of $0\farcs026 \times 0\farcs022$ (P.A. = $-18.8^{\circ}$). The observations were obtained under good atmospheric conditions with precipitable water vapour (PWV) values between 1.2 and 1.8~mm. 

The B7 observations were obtained on 2025 August 10 in a single execution block with an on-source integration time of 490~s. Four FDM spectral windows were centred at 345.75, 347.64, 357.81, and 359.70~GHz, again providing a total continuum bandwidth of 7.5~GHz. The projected baselines ranged from 67~m to 15.24~km, yielding a synthesised beam of $0\farcs021 \times 0\farcs018$ (P.A. = $-56.3^{\circ}$). Excellent observing conditions prevailed during the observations, with a mean PWV of only 0.4~mm, enabling the highest angular resolution of the dataset. 

The quasar J1603$-$4904 served as the phase calibrator for both observing bands. In B6, J1535$-$4730 was used as the check source, whereas J1604$-$4441 fulfilled this role for the B7 observations. All datasets were calibrated using the standard ALMA pipeline implemented in CASA~6.6.1.17 (Pipeline version 2024.1.0.8). Both observing setups successfully passed the ALMA Quality Assurance Level~2 (QA2) process. Although the B6 QA2 report notes decorrelation in one execution of the check source, additional verification demonstrated that this effect was confined to the check source and did not affect the science target or the final imaging products.

We tested different weightings and decided to use \texttt{robust=-0.5} in the B6 image and natural weighting in the B7 image in order to reach similar resolution in both bands. We give the final angular resolutions in Table~\ref{table:alma_obs}. 

Phase-only self-calibration was explored for both the B6 and B7 datasets in an attempt to improve the fidelity and dynamic range of the image. However, the signal-to-noise ratio of the compact continuum emission was insufficient to derive reliable antenna-based gain solutions on the required solution intervals. As a result, the self-calibration process did not produce a significant improvement in the image quality.

The complementary B6 and B7 observations provide continuum images with angular resolutions of $\sim$20--25~mas, corresponding to spatial scales of 50--60\,AU at the distance of \textit{Apep} (assumed to be 2.4~kpc, but see Sect.~\ref{sec:discussion}). Such resolution is sufficient to resolve the compact central emission and directly measure the projected separation of the colliding-wind binary, previously inferred only indirectly to be approximately 47~mas \citep{Han2020}. Furthermore, the two observing bands enable a characterisation of the mm spectral properties of the compact stellar components while simultaneously probing the morphology of the ionised stellar winds at unprecedented spatial resolution.

To complement the SED analysis in Sect.~\ref{sec:broadband_sed}, we obtained integrated flux densities of the whole system from archival, low-resolution (several arcsec) ALMA observations with the 12-m array in Band 3 (B3) and the 7-m array in B6 and B7 (Project: 2024.1.01244.S, PI: J.~Callingham). For these data, we used the pipeline images available in the ALMA archive \footnote{\url{https://almascience.nrao.edu}}. For all bands, we measured flux densities per spectral window to obtain further spectral information. The archival radio observations of this system at $\sim$0.8--20\,GHz also have a resolution of several arcsec \citep{Bloot2022}, with the exception of the high-resolution study by \cite{Marcote2021} with the Long Baseline Array (LBA). We also retrieved flux densities from the Rapid ASKAP Continuum Survey (RACS) survey at 0.887\,GHz, 1.37\,GHz, and 1.65\,GHz \citep{Hale2021_RACS-low, Duchesne2023_RACS-mid}.

In Table~\ref{table:alma_obs}, we summarise the main details of the ALMA observations. All reported errors are statistical only; for fitting purposes, we add in quadrature an additional 10\% systematic uncertainty from absolute calibration errors.

\begin{table*}[h]
\caption{Summary of the ALMA observations in Bands 6 and 7 analysed in this work. The low resolution observations were retrieved from the ALMA archive.}        
\label{table:alma_obs}    
\centering                       
\begin{tabular}{l c c c c c}    
\hline\hline               
\multirow{2}{*}{Start date} & Integration time & Central frequency & \multirow{2}{*}{Weighting} & Beam size & rms \\      
 & (s) & (GHz) & & (mas$\times$mas) & (\uJybeam) \\
\hline          
2025/07/23 & 943.5   & 236.5  & natural  & 22.9$\times$21.5 & 57 \\
2025/08/10 & 489.9   & 352.7  & $r=-0.5$ & 21.3$\times$16.0 & 57 \\    
\hline       
2015/01/19 & 60.5    & 86.0  & $r=0.5$  & 4000$\times$3200 & 144 \\
2025/11/02 & 6863.5  & 225.5 & $r=0.5$  & 6340$\times$4400 & 370 \\ 
2024/10/25 & 6217.7  & 330.6 & $r=0.5$  & 4510$\times$3040 & 833 \\ 
\hline                                                        
\end{tabular}
\end{table*}

\section{Results} \label{sec:results}

The main result of our analysis of the ALMA observations is the detection of resolved mm emission from the \textit{Apep} system. Below we separate the results from direct measurements in the ALMA images from those derived under certain model assumptions and system parameters.

\subsection{ALMA observations}\label{sec:alma_results}

\begin{figure}
    \centering
    \includegraphics[width=\linewidth]{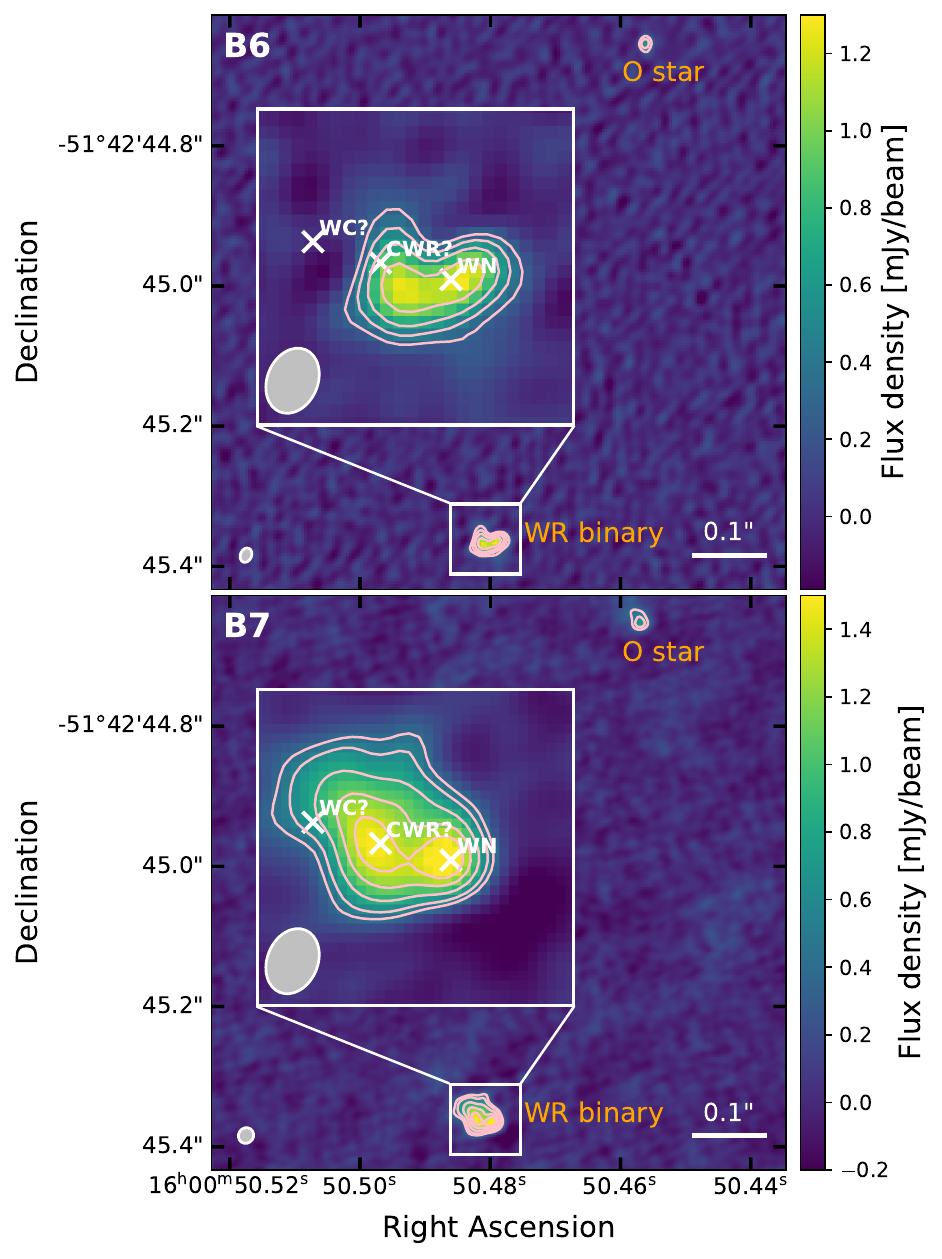} 
    \caption{Continuum images of \textit{Apep} with ALMA at 236.5~GHz (B6, top) and 352.7~GHz (B7, bottom). The pink contour levels are at $n\times rms$, with $n=5,8,12,18$ for the B6 image and $n=7,10,15,20,24$ for the B7 image. The inset highlights the emission from the WR binary. The beam sizes are $22.9\times21.5$\,mas$^2$ for B6 and $21.3\times16.0$\,mas$^2$ for B7.}
    \label{fig:alma_maps}
\end{figure}

\begin{figure}
    \centering
    \includegraphics[width=\linewidth, trim=0 0 0 3, clip]{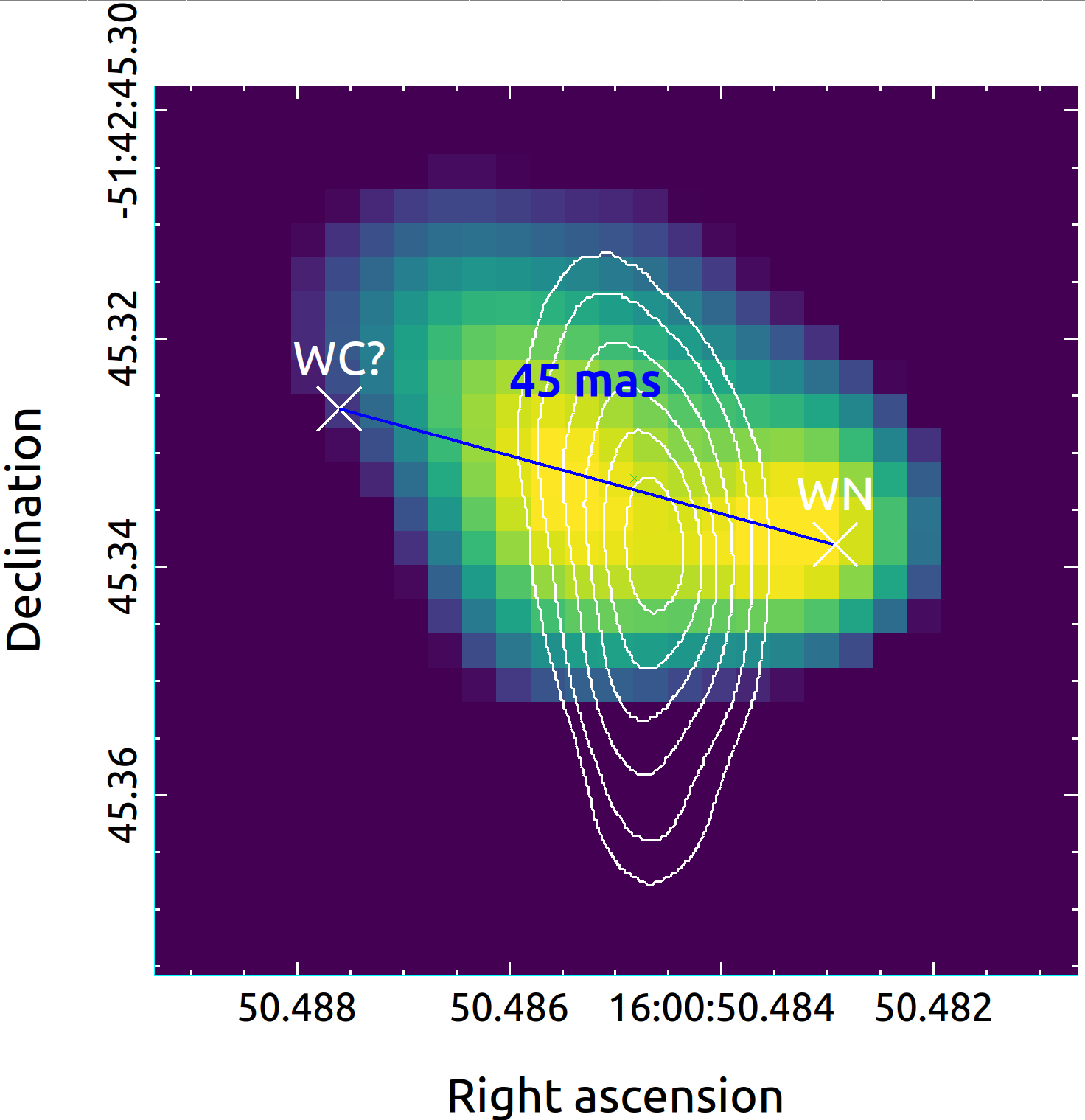} 
    \caption{Natural-weighted ALMA B7 image of \textit{Apep} in colour with white contours from the 2-GHz emission detected with the LBA \citep{Marcote2021}. Contour levels are 0.3, 0.93, 2.81, 5.95, 10.35 and 16\mJybeam. The position of the WC and WN stars are the same as in Fig.~\ref{fig:alma_maps}.}
    \label{fig:lba_vs_alma}
\end{figure}

We present the ALMA continuum images in B6 and B7 in Fig.~\ref{fig:alma_maps}. Qualitatively, the observations reveal structured emission coming from the WR binary and a point-like source at the position of the O star 0.7" away. Within the binary, we find a point-like source consistent with the expected position of the WN star, together with extended emission most likely associated with the CWR (see also Fig.~\ref{fig:lba_vs_alma}). 
The WC star is not clearly identified in the image, suggesting that the emission from its stellar wind is below the sensitivity level of the images. We note that the confident identification of the West component with the WN star is based on the shape of the CWR imaged by \cite{Marcote2021} and the analysis of \cite{delPalacio2022}. 

On a quantitative level, in the B6 image there are two peaks of emission at a level of $\approx$1\mJybeam and $\approx$1.6\mJybeam (the brightest corresponding to the WN star in Fig.~\ref{fig:alma_maps}) separated by an angular distance of $\approx$22\,mas. This separation is consistent with the angular separation between the stars and the CWR, considering the estimated angular separation between the stars ($D_\mathrm{proj} = 47 \pm 6$\,mas; \citealp{Han2020}) and the wind-momentum rates ratio ($\eta = 0.44 \pm 0.08$; \citealp{Marcote2021}), which leads to an expected separation between the WN and the CWR of $D_\mathrm{proj} = D / (1+\sqrt{\eta}) = 28\pm4$\,mas. 

In the B7 image, a point-like source identified as the WN star has a peak flux density of $\approx$1.6\,mJy. The point-like nature of this source has been corroborated using superuniform weighting to increase the angular resolution (at the expense of sensitivity to the fainter structures). The non-detection of the WC star implies an upper limit of $\approx$0.8\mJybeam.

It is not possible to robustly isolate the flux from the structures seen in the images, so we report the total flux of the \textit{Apep} WR binary recovered in both bands. In B6, we measure a flux density ranging between 4--4.8\,mJy depending on the weighting used, with a representative value of $4.5 \pm 0.4$\,mJy from the image in Fig.~\ref{fig:alma_maps}. Similarly, in B7, the total flux density ranges between 5.5 and 7~mJy. 

In addition, the O star presents a flux density of $0.60 \pm 0.09$\,mJy in B6 and $0.9\pm0.4$\,mJy in B7. The spectral index derived from these values is $\alpha = 1.1 \pm 0.9$, which is consistent with the canonical value of $\sim$0.6 within errors.

In a final note, the flux densities obtained from the low resolution Atacama Compact Array (ACA; also known as the Morita Array) observations (which add up the inner binary and the O star) are $S_\mathrm{225\,GHz} = 6.9\pm0.5$\,mJy and $S_\mathrm{331\,GHz} = 6.5\pm0.4$\,mJy. This implies that most of the mm flux is indeed produced on small angular scales around the inner binary, and that most of the flux has been recovered in the high resolution images. Regarding the low resolution B3 data, we measured an average flux density at 86~GHz of $10.29\pm0.12$\,mJy with a poorly constrained spectral index of $-1.3\pm1.4$ (measured in the frequency range 84.5--87.5\,GHz), which is consistent with a significant synchrotron contribution at those frequencies, as discussed in Sect.~\ref{sec:broadband_sed}.

To help with the interpretation of the results, we present a complementary composite image showing the ALMA and LBA 2-GHz data from \cite{Marcote2021} in Fig.~\ref{fig:lba_vs_alma}. In the LBA image, only the CWR is detected (Sect.~\ref{sec:stellar_winds}). There is an offset of $\sim$9\,mas between the peak of the LBA contours and where the putative CWR peaks in the B7 image. 

\subsection{Stellar wind modelling} \label{sec:stellar_winds}

The two key parameters that characterise the stellar winds are the mass-loss rate $\dot{M}$ and the wind terminal velocity $v$. In \cite{Callingham2020}, the terminal wind velocities of the WC and WN stars were measured to be $v_\mathrm{WC}=2100 \pm 200$\,\kms and $v_\mathrm{WN} = 3500 \pm 100$\,\kms. In addition, the shape of the colliding-wind region imaged in radio by \cite{Marcote2021} constrained the wind momentum rates ratio $\eta = (\dot{M}_\mathrm{WC} v_\mathrm{WC})/(\dot{M}_\mathrm{WN} v_\mathrm{WN}) = 0.44\pm0.08$. Using these values and a detailed modelling of the radio SED, \cite{delPalacio2022} estimated mass-loss rates of $\dot{M}_\mathrm{WN} \sim 3 \times 10^{-5}$\,\Msunyr and $\dot{M}_\mathrm{WC} \sim 2 \times 10^{-5}$\,\Msunyr, respectively. More recently, \cite{Zhekov2025} revisited the stellar wind parameters (of both the WR stars and the O star) by fitting high-resolution spectroscopic observations in X-rays with \Ch. The values derived by \cite{Zhekov2025} are broadly consistent with previous estimates, namely $\dot{M}_\mathrm{WN} \sim 4.5 \times 10^{-5}$\,\Msunyr and $\dot{M}_\mathrm{WC} \sim 1.3 \times 10^{-5}$\,\Msunyr, while they also provided values for the O star wind, $v_\mathrm{O} \approx 876$\,\kms (significantly lower than the value of 1280\,\kms reported by \citealp{Callingham2020}) and $\dot{M}_\mathrm{O} \approx 1.6 \times 10^{-6}$\,\Msunyr. There is, however, some tension due to the larger value of $\dot{M}_\mathrm{WN}$, which leads to $\eta=0.21\pm0.03$, significantly lower than the value inferred by \cite{Marcote2021}. In all cases, a distance to the system of $d=2.4$\,kpc was adopted \citep{Callingham2019}. 

The canonical model for free--free emission in spherically symmetric ionised stellar winds predicts a flux density \citep[e.g.][]{Wright1975}:
\begin{equation}
    S_{\nu} = \frac{23.2}{d^2} \gamma^{2/3} \left(\frac{Z}{\mu} \frac{\dot{M}_\mathrm{eff}}{v_\infty} \right)^{4/3} g^{2/3} \nu^{2/3}~\mathrm{Jy},
\end{equation} 
where $d$ is in kpc, $\gamma$ is the ratio of electron to ion number density, $\mu$ is the mean atomic weight, $Z$ is the rms ionic charge, $\dot{M}_\mathrm{eff} = \dot{M}/\sqrt{f}$ is the effective mass-loss rate in units of \Msunyr for a clumpy wind with a volume filling factor $f < 1$, $v_\infty$ is the wind terminal velocity in units of \kms, $g(\nu, T)$ is the gaunt factor \citep{vanHoof2014}, which depends only weakly on the wind temperature $T$, and $\nu$ is the observing frequency in Hz. Similar expressions can also be found to estimate the characteristic radio photospheric radius $R_\nu$.

We adopted values similar to those found in the literature for the WR winds \citep{delPalacio2022, Zhekov2025}. In particular, we adopt wind temperatures of $T_\mathrm{w,WN} = 19$\,kK and $T_\mathrm{w,WC} = 18$\,kK, and $\mu_\mathrm{WN}=2$, $\mu_\mathrm{WC}=4$, $\gamma_\mathrm{WN}=\gamma_\mathrm{WC}=1$, and $Z_\mathrm{WN}=Z_\mathrm{WC}=1$. We then infer values of $R_\nu \sim1$\,AU (corresponding to tens of stellar radii) at the observing frequencies of 220--350~GHz for the three stars. This means that even at the high angular resolution of our ALMA observations, the stellar winds act as point-like sources. We also note that at lower frequencies, these sizes are significantly larger. In particular, at 2~GHz they are several tens of AU, which helps to explain why the emission from the stellar winds was filtered out and undetected in the LBA observations from \cite{Marcote2021} with an angular resolution of just a few mas.

Having checked that the stellar wind emission is point-like, we proceed to calculate the expected flux density. Adopting the parameters of \cite{Zhekov2025} and assuming a volume filling factor of $f=0.3$, the flux densities at 235~GHz are $S_\mathrm{WN}=10.5$\,mJy, $S_\mathrm{WC}=1.2$\,mJy, and $S_\mathrm{O}=0.6$\,mJy, and the flux densities at 353~GHz are $S_\mathrm{WN}=13.3$\,mJy, $S_\mathrm{WC}=1.5$\,mJy, and $S_\mathrm{O}=0.8$\,mJy. Thus, the flux densities are above the observed values, most notably for the WN star.

\subsection{Broadband radio SED modelling} \label{sec:broadband_sed}

Here we provide a more comprehensive analysis of the broadband SED from cm to submm frequencies\footnote{The results in this section can be reproduced using the code hosted at \url{https://github.com/santimda/CWB_non-thermal_emission}.}. We follow the same approach as \cite{delPalacio2022}, now including the new data from Sect.~\ref{sec:observations}. The main point to understand the cm-wavelength SEDs is that the WR stellar winds have a combined kinetic power $L_\mathrm{kin} > 10^{38}$~erg\,s$^{-1}$ \citep{delPalacio2022}, which constitutes an abundant energy reservoir to feed emission processes at the CWR. In particular, relativistic particle acceleration leads to the production of synchrotron emission. As it propagates, this emission can be significantly absorbed by the ionised stellar winds, most significantly at frequencies below 2~GHz \citep{delPalacio2022}.

\begin{figure}
    \centering
    \includegraphics[width=\linewidth]{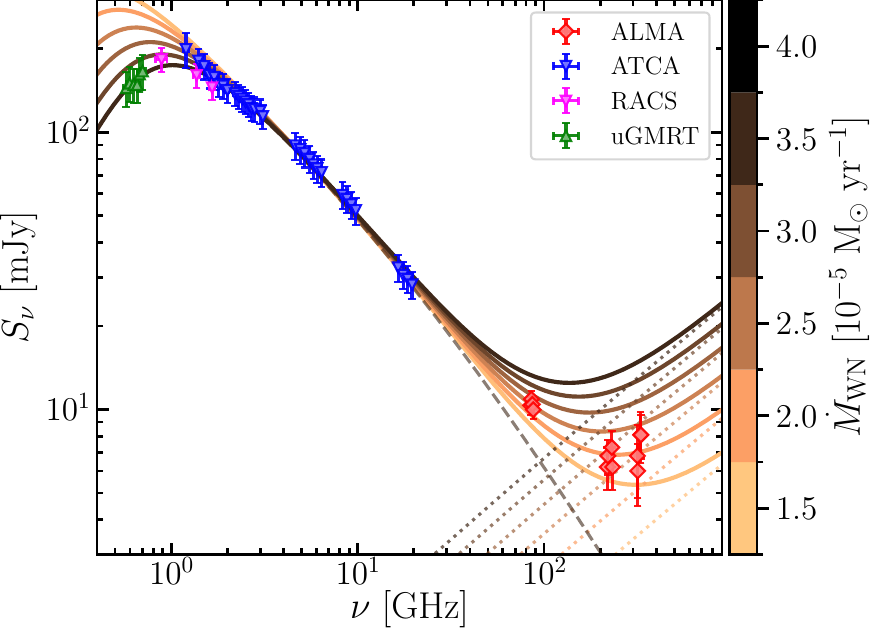} 
    \caption{Broadband cm-to-submm SED of \textit{Apep}. The observational data points are the same as in \cite{delPalacio2022}, with the addition of the low-resolution ALMA points from this work (Sect.~\ref{sec:observations}). We model the SED for different values of the WN star mass-loss rate, as indicated in the colour bar. Dotted lines show the total free--free from the stellar winds, the dashed line the synchrotron component from the CWR, and solid lines are the sum of both.}
    \label{fig:broadband_sed}
\end{figure}

We adopt similar system parameters as \cite{delPalacio2022}, but we update some values to be closer to those recently reported by \cite{Zhekov2025}. Namely, we consider $v_\mathrm{WN}=3350$\,\kms, $v_\mathrm{WC}=2350$\,\kms, and $\eta=0.32$ (the average between those from \citealp{Marcote2021} and \citealp{Zhekov2025}). With this setup, the only free parameter in the model is $\dot{M}_\mathrm{WN}$ \citep{delPalacio2022}. We explore this value systematically and show the results in Fig.~\ref{fig:broadband_sed}. The main conclusion is that values of $\dot{M}_\mathrm{WN} \geq 4 \times 10^{-5}$\,\Msunyr significantly overestimate the free--free flux in the ALMA range and also overestimate the absorption of the low-frequency emission below 1~GHz. Similarly, for $\dot{M}_\mathrm{WN} < 2 \times 10^{-5}$\,\Msunyr, the mm flux drops below the expected values, and the low-frequency absorption is insufficient. 
Finally, a point worth highlighting is that the synchrotron flux from the CWR is $\lesssim$3~mJy at B6 and $\lesssim$2~mJy at B7. Even if most of this flux is produced in the apex of the CWR, it cannot account for all the diffuse flux observed with ALMA (Sect.~\ref{sec:alma_results}); thus, additional emission should come from either free--free radiation in the CWR \citep{Pittard2010radio} or dust. 

\section{Discussion}\label{sec:discussion}

The high-resolution ALMA observations put novel constraints on the emission from the stars in \textit{Apep}. However, the measured flux densities are below the expected ones based on standard wind emission models and previous estimates of the wind properties. Below, we discuss different aspects that can influence this.

First, we address whether variability could affect the observations. The whole dataset used in Sect.~\ref{sec:broadband_sed} spans $\approx$\,8 years, while the orbital period of the inner binary is $193\pm11$~yr \citep{Han2025} and, according to the analysis by \cite{White2025}, the system had its last periastron around 1956. Thus, the system is currently far from periastron and its orbital evolution is very slow. 
Using the orbital solution from \citep{White2025}, we estimate that the change in the stellar separation between the LBA and ALMA B6/B7 observations is $\sim$1\,mas, which is too small to have a significant effect in the radio-to-submm SED. We therefore rule out significant effects due to variability.

Another topic of discussion is the possibly larger distance to the system of $d=4.6^{+0.5}_{-1.8}$~kpc favoured by \cite{Han2025}. When exploring a different value of $d$, one has to keep the constraints imposed by the inferred projected angular separation and the observed radio SED. For fixed wind parameters, assuming a larger $d$ leads to a decrease in the observed wind emission, which at first could seem like a viable alternative to allow for larger $\dot{M}$ values that enhance the free--free absorption while keeping consistency with the ALMA observations. However, a larger $d$ also leads to a larger linear separation $D$ between the stars (given the constraint in $D_\mathrm{proj}$) and thus to a decrease in the free--free opacity for the synchrotron emission, as the emission from the CWR travels on average farther from the stars. The dependence on the physical quantities is such that the increase in $\dot{M}$ required to match the observed ALMA fluxes at a larger $d$ does not enhance the free--free opacity with respect to a lower $\dot{M}$ at a closer distance\footnote{A constant flux needs $\dot{M} \propto d^{3/2}$ (as $S_\nu \propto \dot{M}^{4/3}/d^2$), and as the free--free opacity through the wind scales as $\tau_\mathrm{FFA}\propto \dot{M}^2/D^3$ (with $D \propto d$ for a fixed $D_\mathrm{proj}$), the dependence with $d$ in $\tau_\mathrm{FFA}$ cancels out.}. In other words, the shape of the SEDs shown in Fig.~\ref{fig:broadband_sed} does not change with a larger $d$, only the reference $\dot{M}$ for each curve does. Therefore, the uncertainty in the distance to the system does not help to resolve the tension in the radio-to-submm SED.

There are, however, other sources of uncertainty in the model. For instance, the angular separation between the stars is $47\pm6$~mas. Adopting a smaller separation leads to an increase in the free--free opacity for the same mass-loss rate of the winds, with $D_\mathrm{proj} \approx 40$~mas potentially reconciling the results for $\dot{M}_\mathrm{WN} \approx 2.6\times10^{-5}$\,\Msunyr. However, the non-detection of the WC star in our ALMA observations prevents us from drawing stronger conclusions in this regard. Other sources of uncertainty are the wind clumping (through the parameter $f$, currently fixed at 0.3), although this affects the emission and absorption in the same way. The ionisation structure of the winds can also affect the wind emission, as is $S_{\nu} \propto \gamma^{2/3} \left(\frac{Z}{\mu}\right)^{4/3}$, but the effect should be small if the ratio $Z/\mu$ is not too different from the one adopted in Sect.~\ref{sec:stellar_winds}.

Finally, the fact that the emission of the O star is closer to the expected value than that from the WR stars could indicate that the winds of the WR stars deviate more strongly from the simplified model of \cite{Wright1975}. For instance, other WR stars present continuum emission with a spectral index higher than the canonical $\alpha=0.6$ predicted by this model \citep{Leitherer1991}. One of the strongest assumptions of the model is that of spherically symmetric winds. In the case of \textit{Apep}, other authors have already discussed the possibility of strongly anisotropic winds \citep{Callingham2019, Han2025}. Dedicated modelling efforts are needed to incorporate this effect in both emission and absorption processes in the inner binary to assess whether this can explain the broadband SED of \textit{Apep}. 

\section{Conclusions}

We presented the first high-angular resolution observations at mm wavelengths of \textit{Apep}, the brightest colliding-wind binary at radio wavelengths. The ALMA images revealed the WC star and extended emission in the inner binary but failed to detect the WC component. The O star at 0.7" was also detected. The flux densities measured in B6 and B7 are significantly below the expected values based on published stellar wind parameters and standard emission models of stellar wind; this encouraged us to revisit and discuss some of the assumptions in these models to understand this discrepancy. Moreover, the detection of diffuse emission from the CWR is inconsistent with a simple extrapolation of the synchrotron flux, pinpointing additional emission from either ionised gas or dust. 


\begin{acknowledgements}
We would like to thank the referee for providing useful feedback that helped us to improve the manuscript. S.d.P. thanks S. Zhekov and B. Petrov for insightful discussions about the stellar winds in \textit{Apep}.

This work was carried out in the framework of the PANTERA-Stars\footnote{\url{https://www.astro.uliege.be/~debecker/pantera/}} initiative. This work made use of the software \texttt{ds9} \citep{ds9_2003}, \texttt{CARTA}\footnote{\url{https://cartavis.org/}} and \texttt{Matplotlib} \citep{Matplot_2007}. 

M.F.L. acknowledges support from the PID2023-146675NB-I00 (MCI-AEI-FEDER, UE) program. This work was also partly supported by the Spanish program Unidad de Excelencia María de Maeztu CEX2020-001058-M, financed by MCIN/AEI/10.13039/501100011033, and by the MaX-CSIC Excellence Award MaX4-SOMMA-IC.
V.B-R. acknowledges financial support from the State Agency for Research of the Spanish Ministry of Science and Innovation under grants PID2022-136828NB-C41, PID2025-168247NB-C41 and CEX2024-001451-M funded by MICIU/AEI/10.13039/501100011033/ERDF/EU. V.B-R. is a Correspondent Researcher of CONICET, Argentina, at the IAR.

L.A.Z. acknowledges financial support from CONACyT grant 280775 and UNAM–PAPIIT grants IN110618 and IN112323 (México).

\end{acknowledgements}

\bibliographystyle{aa} 
\bibliography{references} 

@ARTICLE{Benaglia2015,
   author = {{Benaglia}, P. and {Marcote}, B. and {Mold{\'o}n}, J. and {Nelan}, E. and 
	{De Becker}, M. and {Dougherty}, S.~M. and {Koribalski}, B.~S.
	},
    title = "{A radio map of the colliding winds in the very massive binary system HD 93129A}",
  journal = {\aap},
archivePrefix = "arXiv",
   eprint = {1503.07752},
 primaryClass = "astro-ph.SR",
     year = 2015,
    month = jul,
   volume = 579,
      eid = {A99},
    pages = {A99},
      doi = {10.1051/0004-6361/201425595},
   adsurl = {http://adsabs.harvard.edu/abs/2015A%26A...579A..99B}
}

@ARTICLE{DeBecker2013,
   author = {{De Becker}, M. and {Raucq}, F.},
    title = "{Catalogue of particle-accelerating colliding-wind binaries}",
  journal = {\aap},
archivePrefix = "arXiv",
   eprint = {1308.3149},
 primaryClass = "astro-ph.HE",
     year = 2013,
    month = oct,
   volume = 558,
      eid = {A28},
    pages = {A28},
      doi = {},
   adsurl = {https://ui.adsabs.harvard.edu/abs/2013A%26A...558A..28D}
}

@ARTICLE{Eichler1993,
   author = {{Eichler}, D. and {Usov}, V.},
    title = "{Particle acceleration and nonthermal radio emission in binaries of early-type stars}",
  journal = {\apj},
     year = 1993,
    month = jan,
   volume = 402,
    pages = {271-279},
      doi = {},
   adsurl = {http://adsabs.harvard.edu/abs/1993ApJ...402..271E}
}

@ARTICLE{Wright1975,
   author = {{Wright}, A.~E. and {Barlow}, M.~J.},
    title = "{The radio and infrared spectrum of early-type stars undergoing mass loss}",
  journal = {\mnras},
     year = 1975,
    month = jan,
   volume = 170,
    pages = {41-51},
      doi = {10.1093/mnras/170.1.41},
   adsurl = {http://adsabs.harvard.edu/abs/1975MNRAS.170...41W}
}

@ARTICLE{Marcote2015,
       author = {{Marcote}, B. and {Rib{\'o}}, M. and {Paredes}, J.~M. and {Ishwara-Chand
        ra}, C.~H.},
        title = "{Physical properties of the gamma-ray binary LS 5039 through low- and high-frequency radio observations}",
      journal = {\mnras},
         year = "2015",
        month = "Jul",
       volume = {451},
       number = {1},
        pages = {59-73},
          doi = {10.1093/mnras/stv940},
archivePrefix = {arXiv},
       eprint = {1504.07253},
 primaryClass = {astro-ph.HE},
       adsurl = {https://ui.adsabs.harvard.edu/abs/2015MNRAS.451...59M}
}

@ARTICLE{Cappa2004,
       author = {{Cappa}, C. and {Goss}, W.~M. and {van der Hucht}, K.~A.},
        title = "{A Very Large Array 3.6 Centimeter Continuum Survey of Galactic Wolf-Rayet Stars}",
      journal = {\aj},
         year = "2004",
        month = "May",
       volume = {127},
       number = {5},
        pages = {2885-2897},
          doi = {10.1086/383286},
archivePrefix = {arXiv},
       eprint = {astro-ph/0401571},
 primaryClass = {astro-ph},
       adsurl = {https://ui.adsabs.harvard.edu/abs/2004AJ....127.2885C}
}

@ARTICLE{Marcote2021,
       author = {{Marcote}, B. and {Callingham}, J.~R. and {De Becker}, M. and {Edwards}, P.~G. and {Han}, Y. and {Schulz}, R. and {Stevens}, J. and {Tuthill}, P.~G.},
        title = "{AU-scale radio imaging of the wind collision region in the brightest and most luminous non-thermal colliding wind binary Apep}",
      journal = {\mnras},
         year = 2021,
        month = feb,
       volume = {501},
       number = {2},
        pages = {2478-2486},
          doi = {},
archivePrefix = {arXiv},
       eprint = {2012.06571},
 primaryClass = {astro-ph.SR},
       adsurl = {https://ui.adsabs.harvard.edu/abs/2021MNRAS.501.2478M}
}

@ARTICLE{Callingham2020,
       author = {{Callingham}, J.~R. and {Crowther}, P.~A. and {Williams}, P.~M. and {Tuthill}, P.~G. and {Han}, Y. and {Pope}, B.~J.~S. and {Marcote}, B.},
        title = "{Two Wolf-Rayet stars at the heart of colliding-wind binary Apep}",
      journal = {\mnras},
         year = 2020,
        month = jan,
       volume = {495},
       number = {3},
        pages = {3323-3331},
          doi = {},
archivePrefix = {arXiv},
       eprint = {2005.00531},
 primaryClass = {astro-ph.SR},
       adsurl = {https://ui.adsabs.harvard.edu/abs/2020MNRAS.495.3323C}
}

@ARTICLE{Callingham2019,
       author = {{Callingham}, J.~R. and {Tuthill}, P.~G. and {Pope}, B.~J.~S. and {Williams}, P.~M. and {Crowther}, P.~A. and {Edwards}, M. and {Norris}, B. and {Kedziora-Chudczer}, L.},
        title = "{Anisotropic winds in a Wolf-Rayet binary identify a potential gamma-ray burst progenitor}",
      journal = {Nat. Ast.},
         year = 2019,
        month = mar,
       volume = {3},
        pages = {82-87},
          doi = {},
archivePrefix = {arXiv},
       eprint = {1811.06985},
 primaryClass = {astro-ph.SR},
       adsurl = {https://ui.adsabs.harvard.edu/abs/2019NatAs...3...82C}
}

@ARTICLE{Dougherty2003,
       author = {{Dougherty}, S.~M. and {Pittard}, J.~M. and {Kasian}, L. and {Coker}, R.~F. and {Williams}, P.~M. and {Lloyd}, H.~M.},
        title = "{Radio emission models of colliding-wind binary systems}",
      journal = {\aap},
         year = 2003,
        month = oct,
       volume = {409},
        pages = {217-233},
          doi = {10.1051/0004-6361:20031048},
archivePrefix = {arXiv},
       eprint = {astro-ph/0307186},
 primaryClass = {astro-ph},
       adsurl = {https://ui.adsabs.harvard.edu/abs/2003A&A...409..217D}
}

@ARTICLE{Sana2012,
       author = {{Sana}, H. and {de Mink}, S.~E. and {de Koter}, A. and {Langer}, N. and {Evans}, C.~J. and {Gieles}, M. and {Gosset}, E. and {Izzard}, R.~G. and {Le Bouquin}, J. -B. and {Schneider}, F.~R.~N.},
        title = "{Binary Interaction Dominates the Evolution of Massive Stars}",
      journal = {Science},
         year = 2012,
        month = jul,
       volume = {337},
       number = {6093},
        pages = {444},
          doi = {10.1126/science.1223344},
archivePrefix = {arXiv},
       eprint = {1207.6397},
 primaryClass = {astro-ph.SR},
       adsurl = {https://ui.adsabs.harvard.edu/abs/2012Sci...337..444S}
}

@ARTICLE{Puls2008,
       author = {{Puls}, Joachim and {Vink}, Jorick S. and {Najarro}, Francisco},
        title = "{Mass loss from hot massive stars}",
      journal = {\aapr},
         year = 2008,
        month = dec,
       volume = {16},
       number = {3-4},
        pages = {209-325},
          doi = {10.1007/s00159-008-0015-8},
archivePrefix = {arXiv},
       eprint = {0811.0487},
 primaryClass = {astro-ph},
       adsurl = {https://ui.adsabs.harvard.edu/abs/2008A&ARv..16..209P}
}

@ARTICLE{Rosslowe2015,
       author = {{Rosslowe}, C.~K. and {Crowther}, P.~A.},
        title = "{Spatial distribution of Galactic Wolf-Rayet stars and implications for the global population}",
      journal = {\mnras},
         year = 2015,
        month = mar,
       volume = {447},
       number = {3},
        pages = {2322-2347},
          doi = {10.1093/mnras/stu2525},
archivePrefix = {arXiv},
       eprint = {1412.0699},
 primaryClass = {astro-ph.SR},
       adsurl = {https://ui.adsabs.harvard.edu/abs/2015MNRAS.447.2322R}
}

@ARTICLE{Han2020,
       author = {{Han}, Y. and {Tuthill}, P.~G. and {Lau}, R.~M. and {Soulain}, A. and {Callingham}, J.~R. and {Williams}, P.~M. and {Crowther}, P.~A. and {Pope}, B.~J.~S. and {Marcote}, B.},
        title = "{The extreme colliding-wind system Apep: resolved imagery of the central binary and dust plume in the infrared}",
      journal = {\mnras},
         year = 2020,
        month = nov,
       volume = {498},
       number = {4},
        pages = {5604-5619},
          doi = {},
archivePrefix = {arXiv},
       eprint = {2008.05834},
 primaryClass = {astro-ph.SR},
       adsurl = {https://ui.adsabs.harvard.edu/abs/2020MNRAS.498.5604H}
}

@ARTICLE{Bloot2022,
       author = {{Bloot}, S. and {Callingham}, J.~R. and {Marcote}, B.},
        title = "{Radio modelling of the brightest and most luminous non-thermal colliding-wind binary Apep}",
      journal = {\mnras},
         year = 2022,
        month = jan,
       volume = {509},
       number = {1},
        pages = {475-488},
          doi = {},
archivePrefix = {arXiv},
       eprint = {2110.06154},
 primaryClass = {astro-ph.SR},
       adsurl = {https://ui.adsabs.harvard.edu/abs/2022MNRAS.509..475B}
}

@ARTICLE{delPalacio2022,
       author = {{del Palacio}, S. and {Benaglia}, P. and {De Becker}, M. and {Bosch-Ramon}, V. and {Romero}, G.~E.},
        title = "{The non-thermal emission from the colliding-wind binary Apep}",
      journal = {\pasa},
         year = 2022,
        month = jan,
       volume = {39},
          eid = {e004},
        pages = {e004},
          doi = {10.1017/pasa.2021.60},
archivePrefix = {arXiv},
       eprint = {2111.07442},
 primaryClass = {astro-ph.HE},
       adsurl = {https://ui.adsabs.harvard.edu/abs/2022PASA...39....4D}
}

@INPROCEEDINGS{ds9_2003,
       author = {{Joye}, W.~A. and {Mandel}, E.},
        title = "{New Features of SAOImage DS9}",
    booktitle = {Astronomical Data Analysis Software and Systems XII},
         year = 2003,
       editor = {{Payne}, H.~E. and {Jedrzejewski}, R.~I. and {Hook}, R.~N.},
       series = {Astronomical Society of the Pacific Conference Series},
       volume = {295},
        month = jan,
        pages = {489},
       adsurl = {https://ui.adsabs.harvard.edu/abs/2003ASPC..295..489J}
}

@Article{Matplot_2007,
  Author    = {Hunter, J. D.},
  Title     = {Matplotlib: A 2D graphics environment},
  Journal   = {Computing in Science \& Engineering},
  Volume    = {9},
  Number    = {3},
  Pages     = {90--95},
  publisher = {IEEE COMPUTER SOC},
  doi       = {10.1109/MCSE.2007.55},
  year      = 2007
}

@ARTICLE{White2025,
       author = {{White}, Ryan M.~T. and {Pope}, Benjamin J.~S. and {Tuthill}, Peter G. and {Han}, Yinuo and {Dholakia}, Shashank and {Lau}, Ryan M. and {Callingham}, Joseph R. and {Richardson}, Noel D.},
        title = "{The Serpent Eating Its Own Tail: Dust Destruction in the Apep Colliding Wind Nebula}",
      journal = {\apj},
         year = 2025,
        month = nov,
       volume = {994},
       number = {1},
          eid = {121},
        pages = {121},
          doi = {10.3847/1538-4357/adfbe1},
archivePrefix = {arXiv},
       eprint = {2507.14610},
 primaryClass = {astro-ph.SR},
       adsurl = {https://ui.adsabs.harvard.edu/abs/2025ApJ...994..121W}
}

@ARTICLE{Han2025,
       author = {{Han}, Yinuo and {White}, Ryan M.~T. and {Callingham}, Joseph R. and {Lau}, Ryan M. and {Pope}, Benjamin J.~S. and {Richardson}, Noel D. and {Tuthill}, Peter G.},
        title = "{The Formation and Evolution of Dust in the Colliding-wind Binary Apep Revealed by JWST}",
      journal = {\apj},
         year = 2025,
        month = nov,
       volume = {994},
       number = {1},
          eid = {122},
        pages = {122},
          doi = {10.3847/1538-4357/ae12e5},
archivePrefix = {arXiv},
       eprint = {2507.14498},
 primaryClass = {astro-ph.SR},
       adsurl = {https://ui.adsabs.harvard.edu/abs/2025ApJ...994..122H}
}

@INPROCEEDINGS{Hirschi2015,
       author = {{Hirschi}, Raphael},
        title = "{Evolution and Nucleosynthesis of Very Massive Stars}",
    booktitle = {Very Massive Stars in the Local Universe},
         year = 2015,
       editor = {{Vink}, Jorick S.},
       series = {Astrophysics and Space Science Library},
       volume = {412},
        month = jan,
        pages = {157},
          doi = {10.1007/978-3-319-09596-7\_6},
archivePrefix = {arXiv},
       eprint = {1409.7053},
 primaryClass = {astro-ph.SR},
       adsurl = {https://ui.adsabs.harvard.edu/abs/2015ASSL..412..157H}
}

@ARTICLE{vanMarle2010,
       author = {{van Marle}, Allard Jan and {Smith}, Nathan and {Owocki}, Stanley P. and {van Veelen}, Bob},
        title = "{Numerical models of collisions between core-collapse supernovae and circumstellar shells}",
      journal = {\mnras},
         year = 2010,
        month = oct,
       volume = {407},
       number = {4},
        pages = {2305-2327},
          doi = {10.1111/j.1365-2966.2010.16851.x},
archivePrefix = {arXiv},
       eprint = {1004.2791},
 primaryClass = {astro-ph.HE},
       adsurl = {https://ui.adsabs.harvard.edu/abs/2010MNRAS.407.2305V}
}

@INPROCEEDINGS{Offner2023,
       author = {{Offner}, S.~S.~R. and {Moe}, M. and {Kratter}, K.~M. and {Sadavoy}, S.~I. and {Jensen}, E.~L.~N. and {Tobin}, J.~J.},
        title = "{The Origin and Evolution of Multiple Star Systems}",
    booktitle = {Protostars and Planets VII},
         year = 2023,
       editor = {{Inutsuka}, S. and {Aikawa}, Y. and {Muto}, T. and {Tomida}, K. and {Tamura}, M.},
       series = {Astronomical Society of the Pacific Conference Series},
       volume = {534},
        month = jul,
        pages = {275},
          doi = {10.48550/arXiv.2203.10066},
archivePrefix = {arXiv},
       eprint = {2203.10066},
 primaryClass = {astro-ph.SR},
       adsurl = {https://ui.adsabs.harvard.edu/abs/2023ASPC..534..275O}
}

@ARTICLE{Panagia1975,
       author = {{Panagia}, N. and {Felli}, M.},
        title = "{The spectrum of the free-free radiation from extended envelopes.}",
      journal = {\aap},
         year = 1975,
        month = feb,
       volume = {39},
        pages = {1-5},
       adsurl = {https://ui.adsabs.harvard.edu/abs/1975A&A....39....1P}
}

@ARTICLE{Leitherer1991,
       author = {{Leitherer}, Claus and {Robert}, Carmelle},
        title = "{Observations of Stellar Winds from Hot Stars at 1.3 Millimeters}",
      journal = {\apj},
         year = 1991,
        month = aug,
       volume = {377},
        pages = {629},
          doi = {10.1086/170390},
       adsurl = {https://ui.adsabs.harvard.edu/abs/1991ApJ...377..629L}
}

@ARTICLE{Daley-Yates2016,
       author = {{Daley-Yates}, S. and {Stevens}, I.~R. and {Crossland}, T.~D.},
        title = "{Submillimetre free-free emission from the winds of massive stars in the age of Atacama Large Millimeter/submillimeter Array}",
      journal = {\mnras},
         year = 2016,
        month = dec,
       volume = {463},
       number = {3},
        pages = {2735-2745},
          doi = {10.1093/mnras/stw2184},
archivePrefix = {arXiv},
       eprint = {1608.08380},
 primaryClass = {astro-ph.SR},
       adsurl = {https://ui.adsabs.harvard.edu/abs/2016MNRAS.463.2735D}
}

@ARTICLE{Pittard2006radio,
       author = {{Pittard}, J.~M. and {Dougherty}, S.~M. and {Coker}, R.~F. and {O'Connor}, E. and {Bolingbroke}, N.~J.},
        title = "{Radio emission models of colliding-wind binary systems. Inclusion of IC cooling}",
      journal = {\aap},
         year = 2006,
        month = feb,
       volume = {446},
       number = {3},
        pages = {1001-1019},
          doi = {10.1051/0004-6361:20053649},
archivePrefix = {arXiv},
       eprint = {astro-ph/0510283},
 primaryClass = {astro-ph},
       adsurl = {https://ui.adsabs.harvard.edu/abs/2006A&A...446.1001P}
}

@ARTICLE{Pittard2010radio,
       author = {{Pittard}, J.~M.},
        title = "{3D models of radiatively driven colliding winds in massive O + O star binaries - II. Thermal radio to submillimetre emission}",
      journal = {\mnras},
         year = 2010,
        month = apr,
       volume = {403},
       number = {4},
        pages = {1633-1656},
          doi = {10.1111/j.1365-2966.2010.15516.x},
archivePrefix = {arXiv},
       eprint = {0908.1003},
 primaryClass = {astro-ph.SR},
       adsurl = {https://ui.adsabs.harvard.edu/abs/2010MNRAS.403.1633P}
}

@ARTICLE{delPalacio2023,
       author = {{del Palacio}, S. and {Garc{\'\i}a}, F. and {De Becker}, M. and {Altamirano}, D. and {Bosch-Ramon}, V. and {Benaglia}, P. and {Marcote}, B. and {Romero}, G.~E.},
        title = "{Evidence for non-thermal X-ray emission from the double Wolf-Rayet colliding-wind binary Apep}",
      journal = {\aap},
         year = 2023,
        month = apr,
       volume = {672},
          eid = {A109},
        pages = {A109},
          doi = {10.1051/0004-6361/202245505},
archivePrefix = {arXiv},
       eprint = {2302.08170},
 primaryClass = {astro-ph.HE},
       adsurl = {https://ui.adsabs.harvard.edu/abs/2023A&A...672A.109D}
}

@INPROCEEDINGS{Ediss04,
       author = {{Ediss}, G.~A. and {Carter}, M. and {Cheng}, J. and {Effland}, J.~E. and {Grammer}, W. and {Horner}, Jr., N. and {Kerr}, A.~R. and {Koller}, D. and {Lauria}, E.~F. and {Morris}, G. and {Pan}, S.-K. and {Reiland}, G. and {Sullivan}, M.},
        title = "{ALMA Band 6 Cartridge: Design and Performance}",
    booktitle = {Fifteenth International Symposium on Space Terahertz Technology},
         year = 2004,
       editor = {{Narayanan}, Gopal},
        month = jan,
        pages = {181-188},
       adsurl = {https://ui.adsabs.harvard.edu/abs/2004stt..conf..181E}
}

@INPROCEEDINGS{Kerr2004,
       author = {{Kerr}, A.~R. and {Pan}, S.-K. and {Lauria}, E.~F. and {Lichtenberger}, A.~W. and {Zhang}, J. and {Pospieszalski}, M.~W. and {Horner}, N. and {Ediss}, G.~A. and {Effland}, J.~E. and {Groves}, R.~L.},
        title = "{The ALMA Band 6 (211-275 GHz) Sideband-Separating SIS Mixer-Preamp}",
    booktitle = {Fifteenth International Symposium on Space Terahertz Technology},
         year = 2004,
       editor = {{Narayanan}, Gopal},
        month = jan,
        pages = {55-61},
       adsurl = {https://ui.adsabs.harvard.edu/abs/2004stt..conf...55K}
}

@ARTICLE{Langer2012,
       author = {{Langer}, N.},
        title = "{Presupernova Evolution of Massive Single and Binary Stars}",
      journal = {\araa},
         year = 2012,
        month = sep,
       volume = {50},
        pages = {107-164},
          doi = {10.1146/annurev-astro-081811-125534},
archivePrefix = {arXiv},
       eprint = {1206.5443},
 primaryClass = {astro-ph.SR},
       adsurl = {https://ui.adsabs.harvard.edu/abs/2012ARA&A..50..107L}
}

@ARTICLE{Nugis1998,
       author = {{Nugis}, T. and {Crowther}, P.~A. and {Willis}, A.~J.},
        title = "{Clumping-corrected mass-loss rates of Wolf-Rayet stars}",
      journal = {\aap},
         year = 1998,
        month = may,
       volume = {333},
        pages = {956-969},
       adsurl = {https://ui.adsabs.harvard.edu/abs/1998A&A...333..956N}
}

@ARTICLE{Zhekov2025,
       author = {{Zhekov}, Svetozar A. and {Petrov}, Blagovest V.},
        title = "{Chandra high-resolution spectra of Apep Plume: Exploring the colliding-wind picture in X-rays}",
      journal = {\aap},
         year = 2025,
        month = jan,
       volume = {693},
          eid = {A266},
        pages = {A266},
          doi = {10.1051/0004-6361/202452599},
       adsurl = {https://ui.adsabs.harvard.edu/abs/2025A&A...693A.266Z}
}

@ARTICLE{Duchesne2023_RACS-mid,
       author = {{Duchesne}, S.~W. and {Thomson}, A.~J.~M. and {Pritchard}, J. and {Lenc}, E. and {Moss}, V.~A. and {McConnell}, D. and {Wieringa}, M.~H. and {Whiting}, M.~T. and {Wang}, Z. and {Wang}, Y. and {Rose}, K. and {Raja}, W. and {Murphy}, Tara and {Leung}, J.~K. and {Huynh}, M.~T. and {Hotan}, A.~W. and {Hodgson}, T. and {Heald}, G.~H.},
        title = "{The Rapid ASKAP Continuum Survey IV: continuum imaging at 1367.5 MHz and the first data release of RACS-mid}",
      journal = {\pasa},
         year = 2023,
        month = aug,
       volume = {40},
          eid = {e034},
        pages = {e034},
          doi = {10.1017/pasa.2023.31},
archivePrefix = {arXiv},
       eprint = {2306.07194},
 primaryClass = {astro-ph.IM},
       adsurl = {https://ui.adsabs.harvard.edu/abs/2023PASA...40...34D}
}

@ARTICLE{Hale2021_RACS-low,
       author = {{Hale}, Catherine L. and {McConnell}, D. and {Thomson}, A.~J.~M. and {Lenc}, E. and {Heald}, G.~H. and {Hotan}, A.~W. and {Leung}, J.~K. and {Moss}, V.~A. and {Murphy}, T. and {Pritchard}, J. and {Sadler}, E.~M. and {Stewart}, A.~J. and {Whiting}, M.~T.},
        title = "{The Rapid ASKAP Continuum Survey Paper II: First Stokes I Source Catalogue Data Release}",
      journal = {\pasa},
         year = 2021,
        month = nov,
       volume = {38},
          eid = {e058},
        pages = {e058},
          doi = {10.1017/pasa.2021.47},
archivePrefix = {arXiv},
       eprint = {2109.00956},
 primaryClass = {astro-ph.GA},
       adsurl = {https://ui.adsabs.harvard.edu/abs/2021PASA...38...58H}
}

@ARTICLE{vanHoof2014,
       author = {{van Hoof}, P.~A.~M. and {Williams}, R.~J.~R. and {Volk}, K. and {Chatzikos}, M. and {Ferland}, G.~J. and {Lykins}, M. and {Porter}, R.~L. and {Wang}, Y.},
        title = "{Accurate determination of the free-free Gaunt factor - I. Non-relativistic Gaunt factors}",
      journal = {\mnras},
         year = 2014,
        month = oct,
       volume = {444},
       number = {1},
        pages = {420-428},
          doi = {10.1093/mnras/stu1438},
archivePrefix = {arXiv},
       eprint = {1407.5048},
 primaryClass = {astro-ph.HE},
       adsurl = {https://ui.adsabs.harvard.edu/abs/2014MNRAS.444..420V}
}

@ARTICLE{Tasseroul2025,
       author = {{Tasseroul}, M. and {De Becker}, M. and {Blanco}, A.~B. and {Benaglia}, P. and {del Palacio}, S.},
        title = "{Foreground and internal free-free absorption in particle-accelerating colliding-wind binaries: Insights from the radio emission of WR147}",
      journal = {\aap},
         year = 2025,
        month = sep,
       volume = {701},
          eid = {A243},
        pages = {A243},
          doi = {10.1051/0004-6361/202555803},
archivePrefix = {arXiv},
       eprint = {2508.10506},
 primaryClass = {astro-ph.SR},
       adsurl = {https://ui.adsabs.harvard.edu/abs/2025A&A...701A.243T}
}


\end{document}